# Influence of correlations in active medium on pump-induced exceptional points and strong coupling

I. S. Pashkevich[1], I. V. Doronin[1,2], A. A. Zyablovsky[1,2,3], E. S. Andrianov[1,2,3]

[1]*Moscow Institute of Physics and Technology, 9 Institutskiy pereulok, Moscow 141700, Russia*

[2]*Dukhov Research Institute of Automatics (VNIIA), 22 Sushchevskaya, Moscow 127055, Russia*

[3]*Institute for Theoretical and Applied Electromagnetics, 13 Izhorskaya, Moscow 125412, Russia*

Exceptional points show great prospects for applications such as imaging, sensing and designing lasers. Recently, systems with amplifying active medium exhibiting pump-induced exceptional points have attracted much attention due to possibility of controlling strong coupling between light and matter with the aid of pumping. In such structures, the interaction of active molecules with external degrees of freedom, such as phonons or impurities, leads to the destruction of the correlation between polarizations of different molecules. We study the effect of the correlations decay on a system behavior near pump-induced exceptional point. We show that strong coupling persists and eigenvectors together with eigenfrequencies coalesce at a negative value of population inversion, regardless of correlation decay magnitude. Thus, we show that exceptional points are robust to correlation decay, which is crucial for designing systems with exceptional points.

**Introduction**

Non-Hermitian systems have attracted much attention due to their unusual properties. One of their most intriguing features is manifestation of exceptional points (EPs). Exceptional points are singularities in the parameter space of a system at which two eigenvectors and their respective eigenvalues coalesce, resulting in non-complete basis of eigenvectors [1-3]. Passing through an EP is commonly accompanied by non-Hermitian phase transition [4]. Because of their abrupt nature, EPs lead to many remarkable phenomena in optics including unidirectional invisibility [5, 6], loss-induced transparency [7], band merging [8], topological energy transfer [9, 10] and enhanced sensitivity [11-14].

EPs can be realized in a multitude of systems, including coupled microcavities [4], lasers [15, 16] and qubits [17-19]. Lasers and some systems of coupled qubits require active medium as an integral part, whereas coupled microcavities can make use of active medium to achieve gain or loss parameters necessary for manifestation of an EP. Therefore, active medium is a useful component for realizing optical EP systems. One prominent example of EP system is single-mode laser provided the coupling strength between active medium and cavity is larger than characteristic relaxation rates in laser [20]. In this case the system exhibits pump-induced EP at a certain negative value of stationary population inversion [20, 21]. Use of EPs enables to control lasing via mode selection [16, 22, 23], suppression and revival of lasing [15], and through promoting gain in otherwise lossy systems [24]. Additionally, modulating pump power in the vicinity of laser EP results in parametric instability and in coherent emission without population inversion [21, 25].

Interactions between individual quantum emitters and formation of correlations between them have profound effect on the behavior of the system, e.g., cross-relaxation between different polaritonic branches redirect pump drive in the system [26]. Spatial correlations have been shown to play a significant role in lasing [27], in Bose-Einstein condensate based polariton lasers [28, 29], superradiant lasers [30, 31], quantum computers [32] and qubit systems [19, 33, 34]. Correlations have been studied in dissipative hybrid system [35] and in exciton-polariton condensates [36]. However, the effect of polarization correlations in active medium on EPs in laser has not been studied so far, which is the scope of this work.

In this paper, we study the effect of correlation between polarizations of different molecules on the EP in laser consisting of active medium coupled to single-mode cavity in strong coupling regime. We develop a model that enables description of active molecules' correlations in the active medium. We show that increase in dissipations of correlations does not cause disappearance of the EP, i.e. two eigenvalues and two corresponding eigenvectors coalesce and do not form a complete basis at a certain negative value of population inversion. The frequency splitting persists at low pump rate even in presence of large correlation decay, and diminishes with increase in the pump rate, thus maintaining strong coupling regime and eigenfrequency coalescence. We believe that this result is crucial for understanding the role of correlations between polarizations of active medium molecules in non-Hermitian systems with EP and designing such systems.

**Methods**

We investigate properties of a laser near EP [37]. For this purpose, we consider active medium consisting of $N$ two-level active molecules. The active medium is placed in a cavity. We assume that only one cavity mode is excited, other modes have large detuning from the active medium transition frequency and do not influence dynamics of the system. For simplicity, we also assume that the entire active medium occupies a subwavelength volume. All of the molecules have the same transition frequency, which is equal to the eigenfrequency of the cavity mode.

We use the Jaynes-Cummings Hamiltonian in the rotating wave approximation [38] to describe the interaction between the active medium and the electromagnetic field in the cavity:

$$\hat{H} = \hbar\omega\hat{a}^\dagger\hat{a} + \sum_j \hbar\omega\hat{\sigma}_j^\dagger\hat{\sigma}_j + \sum_j \hbar\Omega_R(\hat{a}^\dagger\hat{\sigma}_j + \hat{a}\hat{\sigma}_j^\dagger) \qquad (1)$$

Here $\hat{a}^\dagger$ and $\hat{a}$ are the creation and annihilation operators for photons ($[\hat{a},\hat{a}^\dagger]=1$); $\hat{\sigma}_j^\dagger$ and $\hat{\sigma}_j$ are the raising and lowering operators between two levels of jth molecule ($\{\hat{\sigma},\hat{\sigma}^\dagger\}=1$). $\omega$ is the transition frequency of molecules and the eigenfrequency of the cavity mode. $\Omega_R$ is the coupling constant between an each of the active molecules and the cavity mode. We assume that the coupling constant, $\Omega_R$, is identical for all molecules because the active medium occupies a subwavelength volume.

We use Heisenberg equations [39] to derive differential equations on operators $\hat{a}^\dagger\hat{a}$, $\hat{a}^\dagger\hat{\sigma}_j$, $\hat{\sigma}_i^\dagger\hat{\sigma}_j$ ($i \neq j$) and $\hat{D}_j = \hat{\sigma}_j^\dagger\hat{\sigma}_j - \hat{\sigma}_j\hat{\sigma}_j^\dagger$, then we rescale operators as follows $\hat{n} = \frac{\hat{a}^\dagger\hat{a}}{N}$, $\hat{\varphi} = \frac{i}{2\sqrt{N^3}}\left(\hat{a}\sum_j\hat{\sigma}_j^\dagger - \hat{a}^\dagger\sum_j\hat{\sigma}_j\right)$, $\hat{s} = \frac{1}{N(N-1)}\sum_{i,j(i\neq j)}\hat{\sigma}_i^\dagger\hat{\sigma}_j$, $\hat{D} = \frac{1}{N}\sum_j(\hat{\sigma}_j^\dagger\hat{\sigma}_j - \hat{\sigma}_j\hat{\sigma}_j^\dagger)$ and transition to averages. $n = \langle \hat{a}^\dagger \hat{a} \rangle$ is the average value of photons per one molecule; $D = \langle \hat{D} \rangle$ is the average population inversion of the active medium; $\varphi = \langle \hat{\varphi} \rangle$ is the value of the energy flow between molecules and EM field [40] caused by interaction between polarization of one active molecule and the cavity mode, averaged over all molecules and scaled down by a factor of $1/\sqrt{N}$; $s = \langle \hat{s} \rangle$ is the average spatial correlations between polarizations of two molecules [27]. This specific choice of scaling is the most convenient for our purposes, since it equalizes order of magnitude for each variable. To describe relaxation processes we use Lindblad equation for the density matrix of the system (for more detailed derivation, see [27, 41] and Appendix). Thus we get the following equations [27]:

$$\frac{dn}{dt} = -2\gamma_a n + 2\sqrt{N}\Omega_R \varphi \qquad (2)$$

$$\frac{dD}{dt} = \gamma_P(1-D) - \gamma_D(1+D) - 4\sqrt{N}\Omega_R \varphi \qquad (3)$$

$$\frac{d\varphi}{dt} = -(\gamma_\sigma + \gamma_a + \frac{\gamma_{cor}}{4})\varphi + \frac{\Omega_R}{2}(D+1) + \sqrt{N}\Omega_R n D + \frac{N-1}{\sqrt{N}}\Omega_R s \qquad (4)$$

$$\frac{ds}{dt} = -(2\gamma_\sigma + \gamma_{cor})s + 2\sqrt{N}\Omega_R \varphi D \qquad (5)$$

Here $\gamma_a$ is the relaxation rate of the EM field; $\gamma_D$ and $\gamma_\sigma$ are the longitudinal and transverse relaxation rates of molecules, respectively, $\gamma_\sigma = \gamma_{ph} + \frac{\gamma_P}{2} + \frac{\gamma_D}{2}$ (see Appendix). $\gamma_{ph}$ is the dephasing rate; $\gamma_P$ is the pump rate of active molecules. $\gamma_{cor}$ is an additional relaxation rate for spatial correlations [27] (see Appendix for more details). This parameter determines the value of correlation between active molecules in the system. The value $\gamma_{cor}/\gamma_\sigma = 0$ corresponds to the absence of additional relaxations for the correlations between molecular polarizations. For $\gamma_{cor} \gg \gamma_\sigma$ correlations tend to zero, $s \approx 0$, and have nearly no impact on the dynamics of the system. We assume that $\gamma_a = 5\times10^{-5}\omega$, $\gamma_{ph} = 5\times10^{-4}\omega$, $\gamma_D = 2\times10^{-5}\omega$, $\Omega = 1\times10^{-5}\omega$, $N = 10^6$. These values correspond to a cavity with quality factor $Q = 10^4$, volume $V = 1.7\times10^{-10}cm^3$, and active medium consisting of quantum dots with emission wavelength $\lambda \sim 700nm$, dipole moment of transition $d = 60D$ [42], concentration $\rho = 0.6\times10^{16}cm^{-3}$.

We highlight that relaxation terms $\gamma_a$, $\gamma_\sigma$, $\gamma_D$, $\gamma_P$ in Eqns. (2)-(5) are well-known and can be obtained via Lindblad equation [41]. In Appendix we demonstrate Lindblad equation that produces terms with $\gamma_{cor}$ for the case of two active molecules. The crucial property of these terms is that correlation decay is 4 times faster than the decay of energy flow (see Appendix). This represents effects that cause high decay rate of correlations, $s$, but only limited influence on relaxation of individual dipole moments (which in turn cause relaxation of energy flow $\varphi$). We emphasize the importance of considering Lindblad equation, since relaxation rates that are not consistent with Lindblad equation may result in unphysical processes [41, 43].

**Eigensystem of laser system**

To study eigenmodes of the system and to show existence of the EP we consider small deviations from stationary solution of Eqns. (2)-(5) which has the form:

$$n_{st} = -\frac{(1+D_0)(\gamma_{cor}+2\gamma_\sigma)}{2N_{at}D_0(\gamma_{cor}+2\gamma_\sigma+2\gamma_a)} \quad (6)$$

$$\phi_{st} = -\frac{\gamma_a(1+D_0)(\gamma_{cor}+2\gamma_\sigma)}{2D_0 N_{at}^{3/2}\Omega_R(\gamma_{cor}+2\gamma_\sigma+2\gamma_a)} \quad (7)$$

$$s_{st} = -\frac{(1+D_0)\gamma_\sigma}{N_{at}(\gamma_{cor}+2\gamma_a+2\gamma_\sigma)}, \quad (8)$$

where $n_{st}$, $\varphi_{st}$ and $s_{st}$ are stationary values of $n$, $\varphi$ and $s$, respectively. $D_0 = (\gamma_P - \gamma_D)/(\gamma_P + \gamma_D)$ is the stationary value of population inversion in the absence of EM field. To calculate the stationary value of population inversion in the presence of EM field, we obtain a numerical solution of the Eqns. (2)-(5). As illustrated by Figure 1, below EP the value of $D_{st}$ calculated from Eqns. (2)-(5) is very close to stationary value of population inversion in absence of EM field, $D_0 = (\gamma_P - \gamma_D)/(\gamma_P + \gamma_D)$. This enables treating the population inversion as a parameter and reducing the number of variables, similarly to linear analysis of Maxwell-Bloch equations [44].

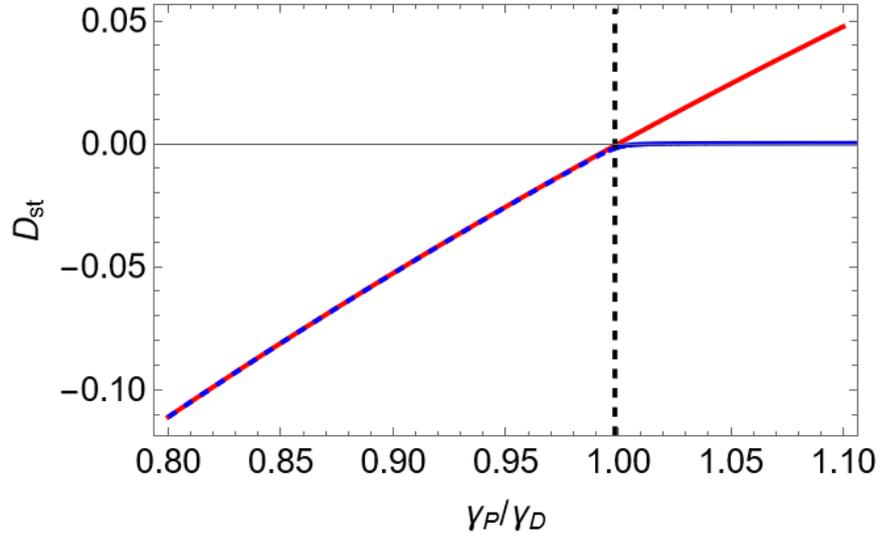

Figure 1. Dependence of the stationary population inversion $D_{st}$, calculated from Eqns. (2)-(5) (the blue line) and stationary population inversion in absence of EM field $D_0$ (the red line) on $\gamma_P/\gamma_D$ at $\gamma_{cor} = 0$. The dashed vertical line denotes EP.

Assuming $n = n_{st} + \delta n$, $\varphi = \varphi_{st} + \delta\varphi$ and $s = s_{st} + \delta s$, we linearize Eqns. (2)-(5) to obtain:

$$\frac{d\mathbf{x}}{dt} = \mathbf{Mx}, \quad (9)$$

where $\mathbf{x} = (\delta n, \delta\varphi, \delta s)^T$, $\delta n$ is a deviation of the average number of photons from the stationary value, $\delta\varphi$ is deviation of the average value of the energy flow between the molecule and field, and $\delta s$ is deviation of the spatial correlations between molecular polarizations. Matrix of the system (9) is

$$\mathbf{M} = \begin{pmatrix} -2\gamma_a & 2\sqrt{N}\Omega_R & 0 \\ \sqrt{N}\Omega_R D_0 & -(\gamma_\sigma + \gamma_a + \dfrac{\gamma_{cor}}{4}) & \dfrac{N-1}{\sqrt{N}}\Omega_R \\ 0 & 2\sqrt{N}\Omega_R D_0 & -(2\gamma_\sigma + \gamma_{cor}) \end{pmatrix} \quad (10)$$

Solving characteristic equation of the system, $\det(\mathbf{M} - \lambda\mathbf{E}) = 0$, we obtain dependencies of eigenvalues $\lambda_{1,2,3}$ on $D_0$ [Figure 2a-f]. It is seen that for $\gamma_{cor} = 0$ there is a point where eigenvalues $\lambda_2$ and $\lambda_3$ are equal to each other suggesting presence of an EP. With increase in $\gamma_{cor}$ these two eigenvalues coalesce at lower values of stationary population inversion $D_0$.

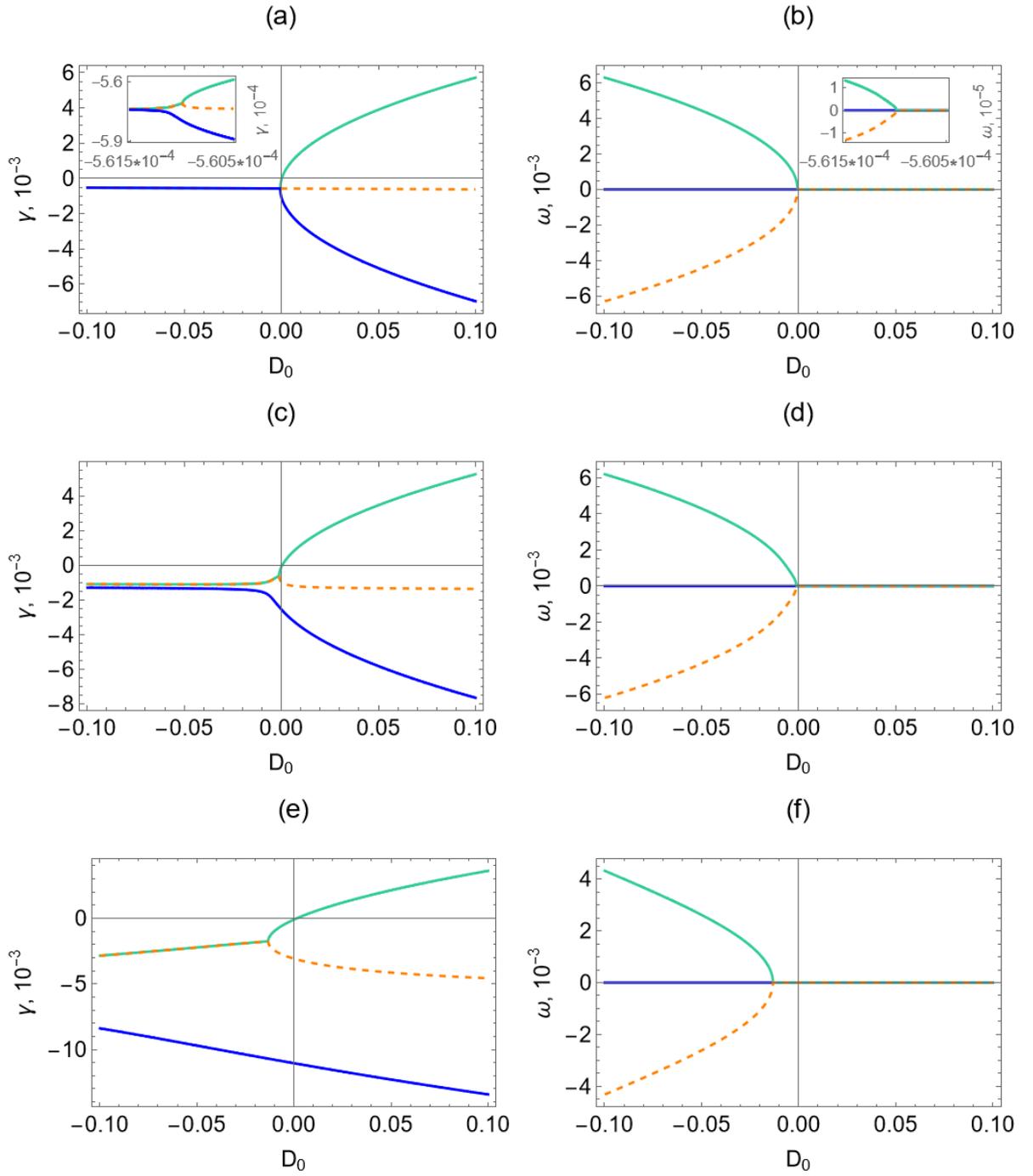

Figure 2. Dependence of the decay rates $\gamma_{1,2,3} = \text{Re}(\lambda_{1,2,3})$ and the oscillation frequencies $\omega_{1,2,3} = \text{Im}(\lambda_{1,2,3})$ ((a, b) - $\gamma_{cor} = 0$, (c, d) - $\gamma_{cor} = 1.5 \times 10^{-3} \omega$, (e, f) - $\gamma_{cor} = 1.0 \times 10^{-2} \omega$) on $D_0$ in Eq. (9). Orange lines represent $\lambda_1$, green lines represent $\lambda_2$ and blue line represent $\lambda_3$. (corresponding to eigenvectors $\mathbf{h}_1$, $\mathbf{h}_2$ and $\mathbf{h}_3$, respectively). Insets in (a), (b) show EP at small scale.

Crucially, coalescence of eigenvalues does not necessarily correspond to an EP [3, 4, 45]. In EP eigenvectors do not form a complete basis. To verify that the system exhibits EP, we find scalar products of eigenvectors of the matrix (10) at different values of additional relaxation rate, $\gamma_{cor}$. The eigenvectors, $\mathbf{h}_j$ are found from the equation $\mathbf{M}\mathbf{h}_j = \lambda_j \mathbf{h}_j$. As seen in Figure 2a-f and Figure 3a-d, the eigenvectors, $\mathbf{h}_2$ and $\mathbf{h}_3$, are linearly dependent

(scalar products of vectors, $\mathbf{h}_2$ and $\mathbf{h}_3$, divided by $|\mathbf{h}_2|$ and $|\mathbf{h}_3|$ is equal to unity) and the eigenvalues $\lambda_2$ and $\lambda_3$ are equal to each other, therefore, eigenvectors $\mathbf{h}_1, \mathbf{h}_2, \mathbf{h}_3$ do not form a complete basis [45]. This proves presence of an EP in the considered system at any additional correlation relaxation.

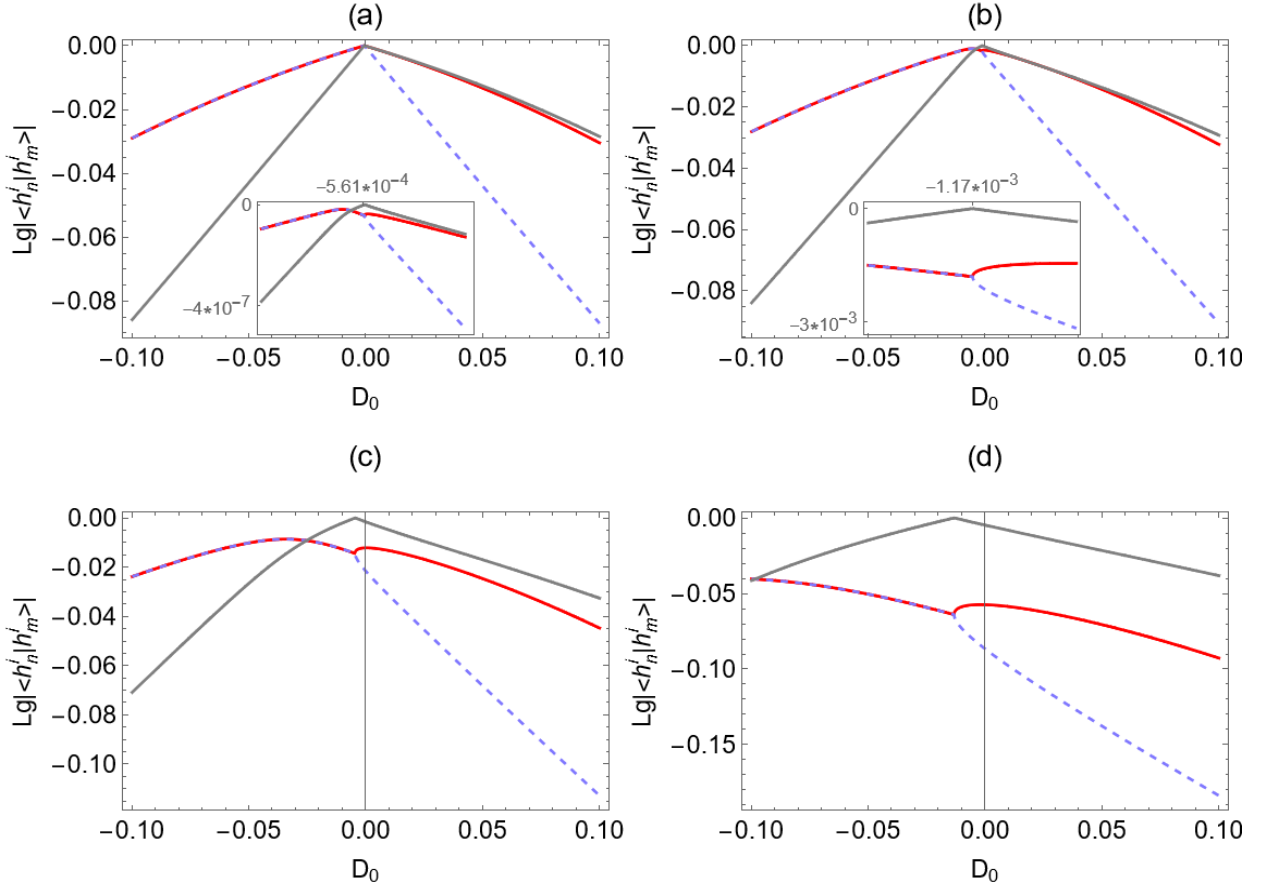

Figure 3. Dependence of absolute value of the scalar products of eigenvectors on $D_0$ for matrix (10) with $\gamma_{cor} = 0$ (a), $\gamma_{cor} = 1.5 \times 10^{-3} \omega$ (b), $\gamma_{cor} = 5.0 \times 10^{-3} \omega$ (c), $\gamma_{cor} = 1.0 \times 10^{-2} \omega$ (d). The red lines correspond to $h_1$ and $h_2$, the purple lines to $h_1$ and $h_3$, the grey lines to $h_2$ and $h_3$. The scalar product of $\mathbf{h}_2$ (orange color on other plots) and $\mathbf{h}_3$ (green color on other plots), divided by $|\mathbf{h}_2|$ and $|\mathbf{h}_3|$ is equal to unity at some value of $D_0$.

## Conclusion

We have studied effect of correlation decay on the parameter space of laser with exceptional point, where two eigenvectors coalesce. The correlations between active molecules play a notable role in the system behavior, causing shifts in eigenfrequencies and in exceptional point position as the correlation decay increases. However, the exceptional point in laser is robust to the correlation decay, persisting even at correlation relaxation rate larger than all other relaxation rates and the coupling strength. The result opens insight in systems with exceptional points. In particular, exceptional points being robust to the correlation decay is crucial to their creation and application.

**Acknowledgements**

The study was financially supported by a Grant from Russian Science Foundation (project No. 20-72-10057).

**Appendix**

Here we detail how Eqns. (2)-(5) on operator averages are obtained from Lindblad equation for the density matrix. Laser is an inherently open system and therefore is subject to relaxation. To describe relaxations in an open system, one must consider the interaction of the system with the environment [41]. In the general case, the Hamiltonian of a system interacting with its environment can be written as

$$\hat{H} = \hat{H}_S + \hat{H}_R + \hat{H}_{SR}, \tag{S1}$$

The first term in (S1) is the Hamiltonian of the isolated system. The second term in (S1) is composed of the Hamiltonians of five separate reservoirs:

$$\hat{H}_R = \hat{H}_R^a + \hat{H}_R^{ph} + \hat{H}_R^D + \hat{H}_R^P + \hat{H}_R^{cor} \tag{S2}$$

All addends in (S2) have general form $\hat{H}_R^i = \sum_k \hbar \omega_k \hat{c}_k^{i\dagger} \hat{c}_k^i$. The operators $\hat{c}_k^i$ and $\hat{c}_k^{i\dagger}$ are the annihilation and creation operators of $k$-th mode of $i$-th reservoir excitation. $\hat{c}_k^i$ and $\hat{c}_k^{i\dagger}$ obey commutation relation, $\left[\hat{c}_k^i, \hat{c}_k^{i\dagger}\right] = 1$, for all reservoirs but pump reservoir, $\hat{H}_R^P$, for which $\left\{\hat{c}_k^P, \hat{c}_k^{P\dagger}\right\} = 1$ instead [41]. The role of each reservoir becomes more apparent after the interaction terms between the system and each reservoir are explicitly written. The Hamiltonian of the interaction of reservoirs with the system has the form:

$$\hat{H}_{SR} = \hat{H}_{SR}^a + \hat{H}_{SR}^{ph} + \hat{H}_{SR}^D + \hat{H}_{SR}^P + \hat{H}_{SR}^{cor} \tag{S3}$$

where $\hat{H}_{SR}^a = \sum_k \hbar \kappa_k^a \hat{a}^\dagger \hat{c}_k^a + h.c.$ describes radiative and non-radiative relaxation of the cavity mode [41], as is evident from presence of operators $\hat{a}$ and $\hat{a}^\dagger$, $\hat{H}_{SR}^{ph} = \sum_j \sum_k \hbar \kappa_k^{ph} \hat{D}_j (\hat{c}_k^{ph} + \hat{c}_k^{ph\dagger})$ describes traverse relaxation in each $j$-th active molecule [41], $\hat{H}_{SR}^D = \sum_j \sum_k \hbar \kappa_k^D \hat{\sigma}_j^\dagger \hat{c}_k^D + h.c.$ describes non-radiative longitudinal relaxation of each $j$-th molecule [41], $\hat{H}_{SR}^P = \sum_j \sum_k \hbar \kappa_k^P \hat{\sigma}_j^\dagger \hat{c}_k^P + h.c.$ describes incoherent pump of each $j$-th molecule [41] (note that $\hat{H}_{SR}^D$ and $\hat{H}_{SR}^P$ seem equivalent, but describe two distinct processes owing to different properties of the corresponding reservoirs), $\hat{H}_{SR}^{cor} = \frac{1}{2} \sum_{i<j} \sum_k \hbar \kappa_k^{cor} (\hat{D}_i - \hat{D}_j) \hat{c}_k^{cor} + h.c.$ describes relaxation for correlations between each $i$-th and $j$-th molecules. As far as we know, the reservoir $\hat{H}_R^{cor}$ with this specific form of interaction $\hat{H}_{SR}^{cor}$ has never been suggested before. Note that the radiative longitudinal relaxation of each molecule is described explicitly via interaction with cavity mode in Eqns. (2)-(5) in the main manuscript. Therefore, reservoir for this process is redundant.

We use the Born-Markov approximation to exclude reservoir variables and obtain the Lindblad equation for the system's density matrix in local approach [46]. Ultimately, this results in the following equation:

$$\frac{\partial \hat{\rho}}{\partial t} = -\frac{i}{\hbar}\left[\hat{H}_S, \hat{\rho}\right] + \hat{L}_a(\hat{\rho}) + \hat{L}_{ph}(\hat{\rho}) + \hat{L}_D(\hat{\rho}) + \hat{L}_P(\hat{\rho}) + \hat{L}_{cor}(\hat{\rho}) \tag{S4}$$

$\hat{L}_i(\hat{\rho})$ are the Lindblad superoperators and share a similar form. $\hat{L}_a(\hat{\rho}) = \gamma_a\left(2\hat{a}\hat{\rho}\hat{a}^\dagger - \hat{a}^\dagger\hat{a}\hat{\rho} - \hat{\rho}\hat{a}^\dagger\hat{a}\right)$ introduces terms $-2\gamma_a n$, and $-\gamma_a \varphi$ to Eqns. (2) and (4). $\hat{L}_{ph}(\hat{\rho}) = 2\gamma_{ph}\left(\hat{D}\hat{\rho}\hat{D} - \hat{\rho}\right)$ introduces terms $-\gamma_{ph}\varphi$ and $-2\gamma_{ph}s$ to Eqns. (4) and (5). $\hat{L}_D(\hat{\rho}) = \gamma_D\left(2\hat{\sigma}\hat{\rho}\hat{\sigma}^\dagger - \hat{\sigma}^\dagger\hat{\sigma}\hat{\rho} - \hat{\rho}\hat{\sigma}^\dagger\hat{\sigma}\right)$ introduces terms $-\gamma_D\varphi/2$, $-\gamma_D s$, $-\gamma_D(1+D)$ to Eqns. (3)-(5). $\hat{L}_P(\hat{\rho}) = \gamma_P\left(2\hat{\sigma}^\dagger\hat{\rho}\hat{\sigma} - \hat{\sigma}\hat{\sigma}^\dagger\hat{\rho} - \hat{\rho}\hat{\sigma}\hat{\sigma}^\dagger\right)$ introduces terms $-\gamma_P\varphi/2$, $-\gamma_P s$, $\gamma_P(1-D)$ to Eqns. (3)-(5). These terms are well-known and are thoroughly covered elsewhere [41]. Finally, most relevant to this work is the relaxation of intermolecular polarizations:

$$\hat{L}_{cor}(\hat{\rho}) = \frac{\gamma_{cor}}{4N}\sum_{i<j}\left((\hat{D}_i - \hat{D}_j)\hat{\rho}(\hat{D}_i - \hat{D}_j) - (1 - \hat{D}_i\hat{D}_j)\hat{\rho} - \hat{\rho}(1 - \hat{D}_i\hat{D}_j)\right) \tag{S5}$$

where indices $i$ and $j$ in the iterate over all active molecules without repeating terms. $\hat{L}_{cor}(\hat{\rho})$ introduces the additional relaxation of correlations to Eqns. (2)-(5). All addends in (S5) collectively add terms $-\gamma_{cor}\varphi(N-1)/(2N)$, $-\gamma_{cor}s$ to equations on $\varphi$ and $s$, respectively. The crucial feature of Eq. (S5) is that the resulting correlation relaxation rate can be as large as four times of the relaxation rate for energy flow, $\varphi$, for $N=2$, unlike all other known form of molecular relaxation listed in Eq. (S4), where this ratio is always equal to two. This enables us to increase relaxation rate for correlations without substantially increasing other relaxation rates. Note, however, that at $N_{at} \to \infty$ the ratio between relaxation rates resulting from Eq. (S5) for $\varphi$ and $s$ tends to 2, and the desired of Eq. (S5) disappears in this limit. However, Eq. (S5) shows that the correlation decay rate can be faster than double of the traverse relaxation rate (as fast as quadruple for Eq. (S5) $N=2$). Addends in Eqns. (4)-(5), $-\gamma_{cor}\varphi/4$, $-\gamma_{cor}s$, are chosen to mimic the properties of relaxation introduced by Eq. (S5) in the best case scenario of $N=2$.

**References**

1. Kato, T., *Perturbation theory for linear operators*. Vol. 132. 2013: Springer Science & Business Media.
2. Heiss, W., *Phases of wave functions and level repulsion.* The European Physical Journal D-Atomic, Molecular, Optical and Plasma Physics, 1999. **7**(1): p. 1-4.
3. Moiseyev, N., *Non-Hermitian quantum mechanics*. 2011: Cambridge University Press.
4. Miri, M.-A. and A. Alù, *Exceptional points in optics and photonics.* Science, 2019. **363**(6422): p. eaar7709.
5. Lin, Z., et al., *Unidirectional invisibility induced by P T-symmetric periodic structures.* Physical Review Letters, 2011. **106**(21): p. 213901.
6. Peng, B., et al., *Parity–time-symmetric whispering-gallery microcavities.* Nature Physics, 2014. **10**(5): p. 394-398.

7. Guo, A., et al., *Observation of P T-symmetry breaking in complex optical potentials.* Physical review letters, 2009. **103**(9): p. 093902.
8. Makris, K.G., et al., *Beam dynamics in P T symmetric optical lattices.* Physical Review Letters, 2008. **100**(10): p. 103904.
9. Doppler, J., et al., *Dynamically encircling an exceptional point for asymmetric mode switching.* Nature, 2016. **537**(7618): p. 76-79.
10. Xu, H., et al., *Topological energy transfer in an optomechanical system with exceptional points.* Nature, 2016. **537**(7618): p. 80-83.
11. Wiersig, J., *Enhancing the sensitivity of frequency and energy splitting detection by using exceptional points: application to microcavity sensors for single-particle detection.* Physical review letters, 2014. **112**(20): p. 203901.
12. Liu, Z.-P., et al., *Metrology with PT-symmetric cavities: enhanced sensitivity near the PT-phase transition.* Physical review letters, 2016. **117**(11): p. 110802.
13. Chen, W., et al., *Exceptional points enhance sensing in an optical microcavity.* Nature, 2017. **548**(7666): p. 192-196.
14. Hodaei, H., et al., *Enhanced sensitivity at higher-order exceptional points.* Nature, 2017. **548**(7666): p. 187-191.
15. Liertzer, M., et al., *Pump-induced exceptional points in lasers.* Physical Review Letters, 2012. **108**(17): p. 173901.
16. Hodaei, H., et al., *Dark-state lasers: mode management using exceptional points.* Optics Letters, 2016. **41**(13): p. 3049-3052.
17. Chen, W., et al., *Decoherence-induced exceptional points in a dissipative superconducting qubit.* Physical Review Letters, 2022. **128**(11): p. 110402.
18. Chen, W., et al., *Quantum jumps in the non-Hermitian dynamics of a superconducting qubit.* Physical Review Letters, 2021. **127**(14): p. 140504.
19. Naghiloo, M., et al., *Quantum state tomography across the exceptional point in a single dissipative qubit.* Nature Physics, 2019. **15**(12): p. 1232-1236.
20. Khurgin, J.B., *Exceptional points in polaritonic cavities and subthreshold Fabry–Perot lasers.* Optica, 2020. **7**(8): p. 1015-1023.
21. Doronin, I., et al., *Lasing without inversion due to parametric instability of the laser near the exceptional point.* Physical Review A, 2019. **100**(2): p. 021801.
22. Hodaei, H., et al., *Parity-time–symmetric microring lasers.* Science, 2014. **346**(6212): p. 975-978.
23. Feng, L., et al., *Single-mode laser by parity-time symmetry breaking.* Science, 2014. **346**(6212): p. 972-975.
24. Zyablovsky, A.A., E.S. Andrianov, and A.A. Pukhov, *Parametric instability of optical non-Hermitian systems near the exceptional point.* Scientific reports, 2016. **6**(1): p. 29709.
25. Doronin, I.V., A.A. Zyablovsky, and E.S. Andrianov, *Strong-coupling-assisted formation of coherent radiation below the lasing threshold.* Optics Express, 2021. **29**(4): p. 5624-5634.
26. Zhang, G.-Q., et al., *Exceptional point and cross-relaxation effect in a hybrid quantum system.* PRX Quantum, 2021. **2**(2): p. 020307.
27. Zyablovsky, A.A., et al., *Long-range atomic correlations as a source of coherent light generation.* Optics Letters, 2021. **46**(21): p. 5292-5295.
28. Byrnes, T., N.Y. Kim, and Y. Yamamoto, *Exciton–polariton condensates.* Nature Physics, 2014. **10**(11): p. 803-813.
29. Deng, H., H. Haug, and Y. Yamamoto, *Exciton-polariton bose-einstein condensation.* Reviews of modern physics, 2010. **82**(2): p. 1489.
30. Meiser, D., et al., *Prospects for a millihertz-linewidth laser.* Physical review letters, 2009. **102**(16): p. 163601.
31. Bohnet, J.G., et al., *A steady-state superradiant laser with less than one intracavity photon.* Nature, 2012. **484**(7392): p. 78-81.
32. Ladd, T.D., et al., *Quantum computers.* nature, 2010. **464**(7285): p. 45-53.
33. Chatterjee, A., et al., *Semiconductor qubits in practice.* Nature Reviews Physics, 2021. **3**(3): p. 157-177.

34. Kjaergaard, M., et al., *Superconducting qubits: Current state of play.* Annual Review of Condensed Matter Physics, 2020. **11**: p. 369-395.
35. Altintas, F. and R. Eryigit, *Dissipative dynamics of quantum correlations in the strong-coupling regime.* Physical Review A, 2013. **87**(2): p. 022124.
36. Hakala, T.K., et al., *Bose–Einstein condensation in a plasmonic lattice.* Nature Physics, 2018. **14**(7): p. 739-744.
37. Choi, Y., et al., *Quasieigenstate coalescence in an atom-cavity quantum composite.* Physical review letters, 2010. **104**(15): p. 153601.
38. Scully, M.O. and M.S. Zubairy, *Quantum optics*. 1997, Cambridge, UK: Cambridge University Press.
39. Landau, L.D. and E.M. Lifshitz, *Quantum Mechanics: Non-Relativistic Theory*. Vol. 3. 1977: Pergamon Press.
40. Zyablovsky, A.A., et al., *Approach for describing spatial dynamics of quantum light-matter interaction in dispersive dissipative media.* Phys. Rev. A, 2017. **95**(5): p. 053835.
41. Carmichael, H.J., *Statistical methods in quantum optics 1: master equations and Fokker-Planck equations*. 2013, New York: Springer Science & Business Media.
42. Guenther, T., et al., *Coherent nonlinear optical response of single quantum dots studied by ultrafast near-field spectroscopy.* Physical review letters, 2002. **89**(5): p. 057401.
43. Shishkov, V.Y., et al., *Relaxation of interacting open quantum systems.* Physics-Uspekhi, 2019. **62**(5): p. 510.
44. Haken, H., *Laser light dynamics*. Vol. 2. 1985: North-Holland Amsterdam.
45. Heiss, W., *The physics of exceptional points.* Journal of Physics A: Mathematical and Theoretical, 2012. **45**(44): p. 444016.
46. Vovchenko, I.V., et al., *Model for the description of the relaxation of quantum-mechanical systems with closely spaced energy levels.* JETP Letters, 2021. **114**: p. 51-57.